\documentclass[aps,prl,amsmath,amssymb,amstext,citeautoscript,punctuation,nofootinbib,superscriptaddress,twocolumn]{revtex4-2}
\usepackage{amsthm,mathrsfs,amsfonts,dsfont,epsfig,braket,enumerate,charter,gensymb,comment,physics,mathtools,natbib}
\usepackage{verbatim} 
\usepackage[normalem]{ulem}
\usepackage{amsmath, amssymb}
\usepackage{bm}
\usepackage{graphicx}
\usepackage{color}
\usepackage{dcolumn}
\usepackage{bm}
\usepackage[hidelinks]{hyperref}
\hypersetup{
	colorlinks   = true, 
	urlcolor     = blue, 
	linkcolor    = blue, 
	citecolor   = red 
}
\usepackage{cleveref}
\usepackage{multirow}

\usepackage[normalem]{ulem}

\begin{document}
	
	\title{Persistent quantum vibronic dynamics in a $5d^1$ double perovskite oxide}
	
	\author{Naoya Iwahara}
	\thanks{These authors contributed equally to this work}
	\email[]{naoya.iwahara@gmail.com}
	\affiliation{Graduate School of Engineering, Chiba University, 1-33 Yayoi-cho, Inage-ku, Chiba-shi, Chiba 263-8522, Japan} 
	\author{Jian-Rui Soh}
	\thanks{These authors contributed equally to this work}
	\email[]{jian.soh@epfl.ch}
	\affiliation{Quantum Innovation Centre (Q.InC), Agency for Science Technology and Research (A*STAR), 2 Fusionopolis Way, Singapore 138634}
	\affiliation{Institute of Physics, \'{E}cole Polytechnique \'{F}e\'{d}erale de Lausanne (EPFL), CH-1015 Lausanne, Switzerland}
	\author{Daigorou Hirai}
	\affiliation{Department of Materials, Physics and Energy Engineering, Nagoya University, Furo-cho, Chikusa-ku, Nagoya, 464-8601, Japan}
	\author{Ivica \v{Z}ivkovi\'c} 
	\affiliation{Institute of Physics, \'{E}cole Polytechnique \'{F}e\'{d}erale de Lausanne (EPFL), CH-1015 Lausanne, Switzerland}
	\author{Yuan Wei}
	\affiliation{Paul Scherrer Institute, Villigen PSI, Switzerland}
	\author{Wenliang Zhang}
	\affiliation{Paul Scherrer Institute, Villigen PSI, Switzerland}
	\author{Carlos Galdino}
	\affiliation{Paul Scherrer Institute, Villigen PSI, Switzerland}
	\author{Tianlun Yu}
	\affiliation{Paul Scherrer Institute, Villigen PSI, Switzerland}
 \author{Kenji Ishii}
	\affiliation{Synchrotron Radiation Research Center, National Institutes for Quantum Science and Technology, Sayo, Hyogo 679-5148, Japan}
	\author{Federico Pisani}
	\affiliation{Institute of Physics, \'{E}cole Polytechnique \'{F}e\'{d}erale de Lausanne (EPFL), CH-1015 Lausanne, Switzerland}
	\author{Oleg Malanyuk}
	\affiliation{Institute of Physics, \'{E}cole Polytechnique \'{F}e\'{d}erale de Lausanne (EPFL), CH-1015 Lausanne, Switzerland}
	\author{Thorsten Schmitt}
	\affiliation{Paul Scherrer Institute, Villigen PSI, Switzerland}
	\author{Henrik M R{\o}nnow}
	\affiliation{Institute of Physics, \'{E}cole Polytechnique \'{F}e\'{d}erale de Lausanne (EPFL), CH-1015 Lausanne, Switzerland}
	
	\date{\today}

	\begin{abstract}
Quantum entanglement between the spin, orbital, and lattice degrees of freedom in condensed matter systems can emerge due to an interplay between spin-orbit and vibronic interactions. Heavy transition metal ions decorated on a face-centered cubic lattice, for example, in $5d^1$ double perovskites, are particularly suited to support these quantum entangled states, but direct evidence has not yet been presented. In this work, we report additional peaks in the low-energy spectra of a $5d^1$ double perovskite, Ba$_2$CaReO$_6$, which cannot be explained by adopting a purely classical description of lattice vibrations. Instead, our theoretical analysis demonstrates that these spectroscopic signatures are characteristic of orbital-lattice entangled states in Ba$_2$CaReO$_6$. Crucially, both theory and experiment demonstrate that these quantum-entangled states persist to low temperatures, despite the onset of multipolar order. \end{abstract}
	
	\maketitle
	
	Within the realm of condensed matter physics, exotic phenomena often trace their origin to the influence of strong quantum effects. For example, in heavy transition metal compounds, the spin-orbit entanglement on metal sites results in diverse phenomena, including Kitaev spin liquid phases, excitonic magnetism, topological insulating phases, and hidden multipolar phases~\cite{Witczak-Krempa2014, Takagi2019, Takayama2021}.
	
	Recent investigations into cubic $5d^1$ double perovskites (DP) highlight the intricate interplay between spin-orbit and lattice degrees of freedom, which govern the multipolar orderings in spin-orbit Mott insulators~\cite{Lu2017, Liu2018, Willa2019, Ishikawa2019, Tehrani2023, Hirai2020, Soh2023, Mosca2021, Merkel2023}. In these materials, the $d^1$ metal sites host four-fold degenerate low-energy electronic states characterized by $j_\text{eff} = 3/2$, which can harbor dipole, quadrupole, and octupole moments. Notably, symmetry allows the time-even quadrupole moments to couple to lattice distortions. As such, the onset of quadrupolar orders in $5d^1$ DPs can be accompanied by the concomitant development of cooperative Jahn-Teller (JT) deformations, i.e. structural order.
	
	This symmetry-allowed coupling finds experimental support in $5d^1$ DP compounds such as Ba$_2$NaOsO$_6$ \cite{Lu2017, Liu2018, Willa2019}, Cs$_2$TaCl$_6$ \cite{Ishikawa2019, Tehrani2023}, and Ba$_2$MgReO$_6$ \cite{Hirai2020, Soh2023} where structural orderings have been observed. A recent study using synchrotron elastic x-ray scattering techniques further confirms these findings by unveiling the simultaneous development of quadrupolar and structural orders~\cite{Soh2023}. Complementing these experimental observations are first-principles calculations, which have successfully replicated quadrupolar orders coupled with cooperative JT deformations \cite{Mosca2021, Merkel2023}.

    However, these previous studies have predominately favored a classical description of the lattice degrees of freedom, which break down for the $5d^1$ DPs. For instance, a resonant inelastic x-ray scattering (RIXS) study of $5d^1$ DPs at the $L_3$ edge finds an asymmetric $j_\text{eff} = 1/2$ peak~\cite{Frontini2023}.
    Crucially, the asymmetric shape of the $j_\text{eff} = 1/2$ peak remains unchanged by structural transitions, thereby ruling out the conventional Franck-Condon mechanism as an explanation.

    The missing piece in many of the former studies is the quantum nature of lattice vibrations which play a decisive role in shaping the low-energy spectra. 
     In particular, the strong spin-orbit coupling in $5d$ ions enhances the quantum entanglement between 
     the $j_\text{eff}=3/2$ ground state on metal sites, with the quantized vibrations of the surrounding lattice  (dynamic JT effect \cite{Bersuker1989}).
     The experimental fingerprints of these orbital-lattice entangled vibronic states are additional transitions, which emerge due to the dressing of the $j_\mathrm{eff}$=3/2
     manifold by lattice vibrations that cannot simply be explained by adopting a purely classical description. 
     Indeed, the transitions between the vibronic ground states and excited $j_\text{eff}=1/2$ states make the $j_\text{eff}=1/2$ RIXS peak of the $5d^1$ DPs asymmetric \cite{Frontini2023}.
    
    Further supporting this notion are first-principles calculations of several $5d^1$ DPs, which indicate that the dynamic JT effect plays a non-negligible role in stabilizing the ground state~\cite{Iwahara2018, Mosca2024}. Moreover, the emergence of multipolar phases observed in the wider family of cubic $5d^1$ DP compounds can trace their natural origin to the ordering of vibronic states at sufficiently low temperatures~\cite{Iwahara2023}.  Nevertheless, several open questions remain unanswered. 
	
	For instance, there is a notable absence of direct spectroscopic evidence for vibronic states in $5d^1$ DPs. This is because the RIXS experiments conducted so far have primarily been focused on the metal absorption $L_3$-edge, resulting in resolution-limited spectra that fail to capture the detailed low-energy features of vibronic states. To conclusively demonstrate the presence of vibronic dynamics, RIXS experiments with higher energy resolution are imperative. One promising avenue is to investigate RIXS at the absorption $K$-edge of the ligands, which are strongly covalently bonded to the metal ions. RIXS at the O $K$-edge is particularly sensitive to vibronic excitations due to its significantly longer core-hole lifetime ($\Gamma$=150 meV)~\cite{johnston_electron-lattice_2016}.
    Indeed, recent O $K$-edge RIXS measurements of $5d^1$ DPs reveal low-lying excitations \cite{Agrestini2024, Zivkovic2024}, but their analyses do not account for the dynamic JT effect.
	
    Furthermore, it is unclear how the JT effect can remain dynamic, even in the presence of static structural order in $5d^1$ DPs. Addressing this question will entail reconciling the dynamic JT effect with static JT distortions. Theoretically, this necessitates an analysis of how the orbital-lattice entanglement evolves as a function of structural distortion. Experimentally, it will require a clear demonstration that the dynamic JT effect can persist across the structural transition temperature.

    In this work, we employ a combined theoretical and experimental approach to address these questions, by investigating the nature of the low-energy quantum states in $5d^1$ DPs. Our theoretical analysis of the evolution of vibronic levels under an external elastic field demonstrates the persistence of orbital-lattice entanglement despite static JT deformations. We directed our attention towards Ba$_2$CaReO$_6$, a $5d^1$ DP system known for exhibiting cubic-to-tetragonal distortion at $T_\mathrm{s}\sim$130\,K~\cite{Yamamura2006, Ishikawa2021}. In order to validate our calculations of the vibronic states above and below $T_\mathrm{s}$, we probed the low-energy states of Ba$_2$CaReO$_6$ using RIXS at both the O $K$- and Re $L_3$-edges. The O $K$-edge spectra, in particular, unveil fine structures that can only be explained in terms of vibronic states, arising from transitions between the ground and low-energy excited vibronic levels. Crucially, our O $K$-RIXS data demonstrate that these vibronic states survive in the structurally ordered phase, in agreement with our calculations, which simultaneously accounts for JT dynamics and structural deformations.
	
	\begin{figure}[t!]
		\centering
		\includegraphics[width=0.49\textwidth]{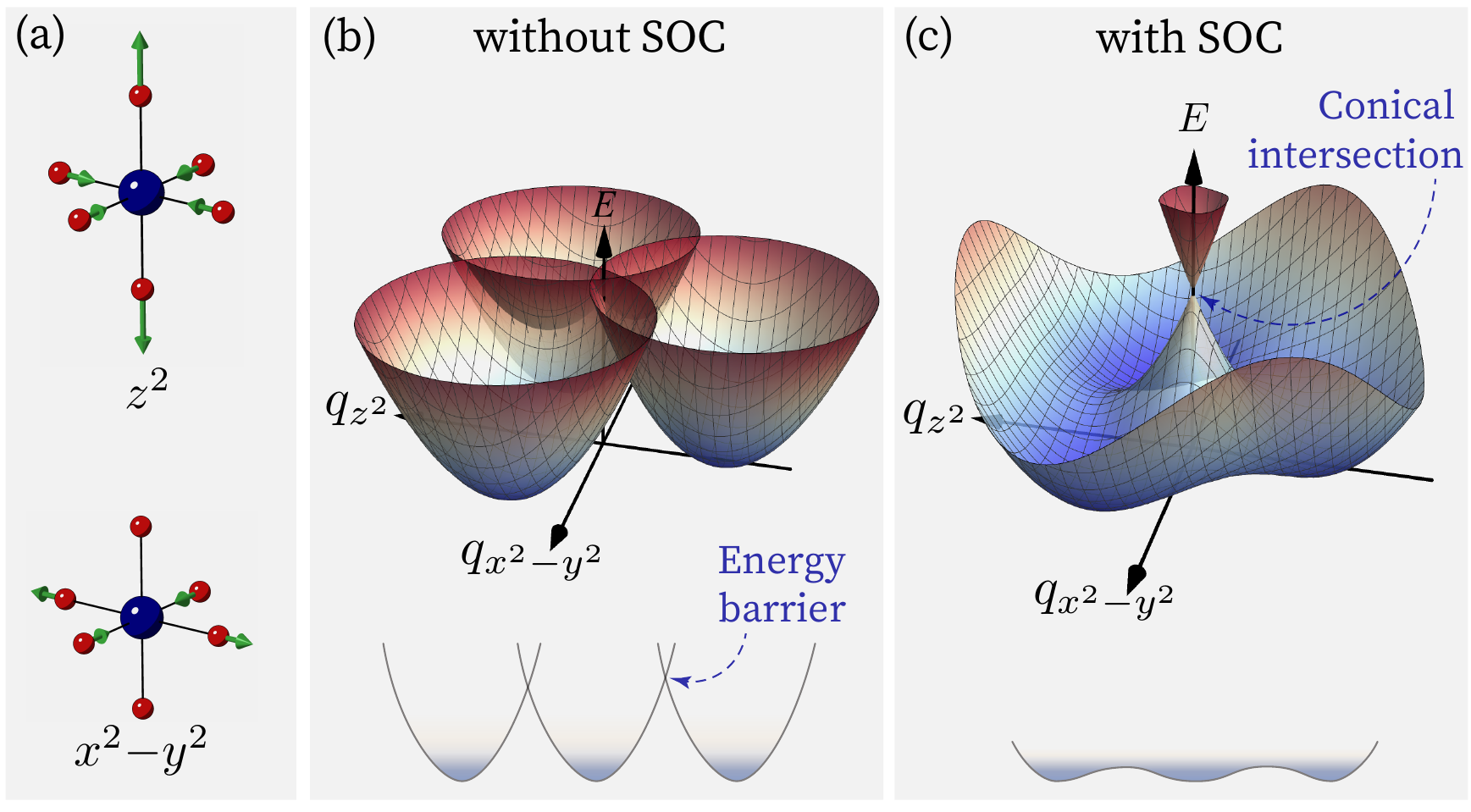}
		\caption{(a) The JT active modes. The APES (b) without and (c) with spin-orbit coupling.}
		\label{Fig:Fig1}
	\end{figure}
		
	To understand how vibronic states arise in the first place, we review the key onsite interactions of a $d^1$ ion residing in an undistorted octahedral environment. Here, the covalency between the metal $d$ and ligand $p$ orbitals splits the five atomic $d$ orbitals into doubly-degenerate $e_g$ and triply-degenerate $t_{2g}$ orbitals \cite{Sugano1970}. Given that the $t_{2g}$ orbitals are lower in energy, they are populated by the single $d$ electron.
	
	On the one hand, due to the spin-orbit interaction, the unquenched orbital angular momentum ($l = 1$) of the $t_{2g}$ orbital can couple linearly to the electron spin~\cite{Sugano1970}, 
	\begin{align}
		\hat{H}_\text{SO} &= \lambda \hat{\bm{l}} \cdot \hat{\bm{s}},
		\label{Eq:HSO}
	\end{align}
	where $\lambda$ is the spin-orbit coupling parameter, and $\hat{\bm{l}}$ and $\hat{\bm{s}}$ are the effective orbital and spin angular momenta, respectively. Correspondingly, the energy eigenstates of $\hat{H}_\text{SO}$ are the $j_\text{eff}=1/2$ $(\Gamma_7)$ and $j_\text{eff}=3/2$ $(\Gamma_8)$ multiplet states with the eigenvalues of $\lambda$ and $-\lambda/2$, respectively. 
	
    On the other hand, by symmetry, these $t_{2g}$ orbitals can also couple vibronically to the JT active modes~\cite{Bersuker1989},
	\begin{align}
		\hat{H}_\text{JT} &= 
		\sum_{\gamma=z^2, x^2-y^2} 
		\frac{ \hslash \omega }{2} \left( \hat{p}_\gamma^2 + \hat{q}_\gamma^2 \right)
		+ \hat{V}_\text{JT},
		\label{Eq:HJT0}
	\end{align}
	where
	\begin{align}
		\hat{V}_\text{JT} &=
		\hslash \omega g ( \hat{q}_{x^2} |yz \rangle \langle yz|
		+ \hat{q}_{y^2} |zx \rangle \langle zx|
		+ \hat{q}_{z^2} |xy \rangle \langle xy|) ,
		\label{Eq:HJT}
	\end{align}
	via $g$, the dimensionless vibronic coupling parameter between the $d$-$p$ hybridized electronic orbitals $|\gamma\rangle$ ($\gamma = yz, zx, xy$) and the lattice distortion modes $\hat{q}_{\gamma}$ ($\gamma$=$x^2$, $y^2$, $z^2$) with vibration frequency of $\omega$. Here, $\hat{p}_\gamma$ is the dimensionless conjugate momenta. Throughout the work, we consider only the $E_g$ vibrational modes [Fig.~\ref{Fig:Fig1}(a)] among the vibronically active modes because their coupling strength is the most dominant 
    in $5d^1$ DPs \cite{Iwahara2018, Frontini2023}. 
	
    The resultant adiabatic potential energy surface (APES) of Eq.~(\ref{Eq:HJT}) -- where the effects of spin-orbit coupling are neglected -- comprises three disconnected harmonic oscillator potentials [Fig.~\ref{Fig:Fig1}(b)]. In such cases, due to the large energy barriers between the degenerate ground states, a purely classical description of the lattice distortion would suffice. However, in cases involving heavy ions (e.g., $5d$ ions) where the effect of spin-orbit interaction becomes important, the potential barrier can be reduced significantly  \cite{Opik1957, Streltsov2020}, which qualitatively changes the quantum states of the system.

	To exemplify this, we have to place the $\hat{H}_\mathrm{SO}$ and $\hat{H}_\mathrm{JT}$ terms on equal footing, and consider the resultant wavefunction. In fact, in an undistorted octahedral environment, symmetry allows for both the spin-orbit and vibronic interactions to coexist. Using the symmetry relations $\Gamma_7 = \Gamma_2 \otimes \Gamma_6$ and $\Gamma_8 = \Gamma_3 \otimes \Gamma_6$, we can decouple the multiplet states into the pseudo-orbital and pseudo-spin parts. The pseudo-orbital parts ($\Gamma_2 \oplus \Gamma_3$) are time-even and couple to the $E_g$ modes,
	\begin{align}
		\hat{V}'_\text{JT} &= 
		\frac{\hslash \omega g}{2}
		\begin{pmatrix} 
			0 & \sqrt{2} \hat{q}_{x^2-y^2} & -\sqrt{2} \hat{q}_{z^2} \\
			\sqrt{2} \hat{q}_{x^2-y^2}     & -\hat{q}_{z^2}     & \hat{q}_{x^2-y^2} \\
			-\sqrt{2} \hat{q}_{z^2} & \hat{q}_{x^2-y^2} & \hat{q}_{z^2} \\
		\end{pmatrix},
		\label{Eq:VJT}
	\end{align}
	with the $\Gamma_2$ and $\Gamma_3$ pseudo-orbital states forming the basis of the matrix. Although the spin-orbit coupling quenches the vibronic coupling in the $j_\text{eff}=3/2$ states by half compared with the case in Eq.~(\ref{Eq:HJT}), the vibronic coupling remains non-zero. 

    Qualitatively, this results in an APES with a shallow barrier between adjacent minima, as shown in Fig.~\ref{Fig:Fig1}(c) \cite{Opik1957, Streltsov2020}. Consequently, the wave function spreads over the circular trough of the APES around the conical intersection.
    The energy eigenstates (vibronic states) of the Hamiltonian $\hat{H} = \hat{H}_\text{SO} + \hat{H}_\text{JT}$ take the orbital-lattice entangled form, 
	\begin{align}
		|\Psi_{\nu \sigma} \rangle &= 
		\Bigg(
		\sum_{\Gamma\gamma}
		|\Gamma\gamma\rangle \otimes |\chi_{\Gamma\gamma;\nu} \rangle 
		\Bigg)
		\otimes |\Gamma_6 \sigma\rangle,
		\label{Eq:vibronic}
	\end{align}
	due to the non-adiabatic effect. 
	Here $|\Gamma\gamma\rangle$, $|\chi\rangle$, and $|\Gamma_6\sigma\rangle$ denote the pseudo-orbital, lattice-vibrational, and pseudo-spin states, respectively. As the vibronic wavefunction traverses along the circular trough around the conical intersection [Fig.~\ref{Fig:Fig1}(c)], the electronic part undergoes a change in sign due to the Berry phase it acquires. This imposes a periodic boundary condition where the corresponding lattice-vibrational part has to also change sign, in order to make $|\Psi_{\nu\sigma}\rangle$ single-valued. The resultant vibronic energy levels follow the sequence $\Gamma_8$, $\Gamma_7$, $\Gamma_6$, $\cdots$ in increasing order of energy~\cite{Ham1987}. The transitions between these vibronic levels give rise to the characteristic spectroscopic fingerprints, which can be detected with RIXS.
	
\begin{figure}[t!]
\centering
\includegraphics[width=0.49\textwidth]{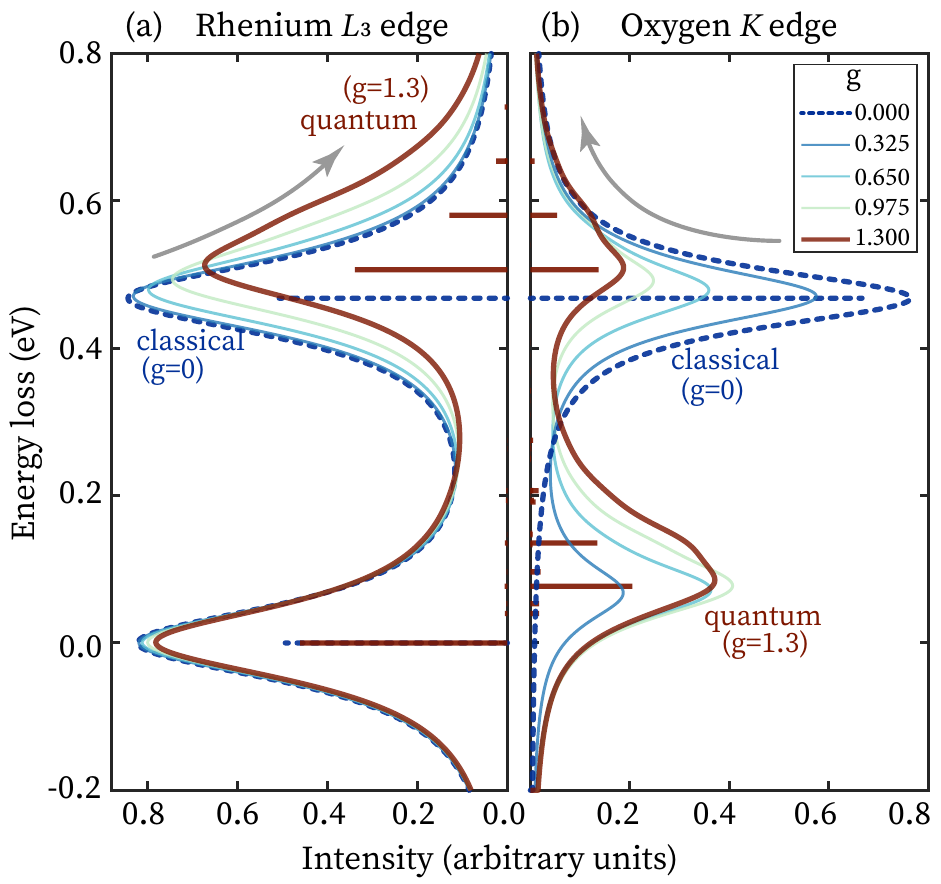}
\caption{The calculated RIXS spectrum at the (a) rhenium $L_3$ and (b) oxygen $K$ edge based on a classical ($g$=0) and quantum ($g$=1.3) description of the lattice degrees of freedom are denoted by the dashed blue and solid red lines, respectively. The faint lines correspond to calculations for intermediate coupling strengths $g$=0.325, 0.650 and 0.975. 
}
\label{Fig:Fig_coupling}
\end{figure}

    To illustrate the spectroscopic signatures of these vibronic transitions, we plot in Fig.~\ref{Fig:Fig_coupling} the calculated RIXS spectra for a range of coupling strengths ($g$), for an undistorted Ba$_2$CaReO$_6$. We employed the Kramers-Heisenberg formula~\cite{Sakurai1967, RIXS} and the dipole approximation to calculate the spectra that arise from transitions between the vibronic states (see computational details).

    In the classical regime ($g$=0), where the orbital and lattice degrees of freedom are decoupled, our calculations reveal a single, purely-electronic $j_\mathrm{eff}$=1/2 transition at $\Delta E$$\sim$0.5 eV. This transition is characterized by a symmetric peak at both the rhenium $L_3$ and oxygen $K$ edges. In contrast, when a quantum description of the lattice is adopted ($g > 0$), the orbital and lattice degrees of freedom become entangled. For instance, with $g$=1.3, the $j_\mathrm{eff}$=1/2 transition broadens asymmetrically and acquires a long tail that extends towards high energies at both the Re $L_3$ and O $K$ edges. This broadening is due to the presence of many additional quantum vibronic transitions, which come from the dressing of the $j_\mathrm{eff}$=3/2 manifold by lattice vibrations. 

    Furthermore, a strong asymmetric peak also appears at $\Delta E$$\sim$ 0.15\,eV of the O $K$ RIXS spectra for 
    $g$=1.3
    [Fig.~\ref{Fig:Fig_coupling}(b)]. The peak originates from a multitude of quantum vibronic transitions between 0-0.4\,eV.
    As shown in Fig.~\ref{Fig:Fig_coupling}(b), such a peak is absent in the classical case, where $g$=0. Therefore, the presence of such a peak is the key spectroscopic fingerprint of quantum vibronic states arising from spin-orbit-lattice entanglement.
    
Vibronic transitions are more pronounced in the RIXS spectra at the O $K$ edge than at the Re $L_3$ edge. This arises because the $K$-edge scattering amplitude contains both magnetic and quadrupolar contributions~\cite{Kim2017}, whereas the $L_3$-edge amplitude is purely quadrupolar—without spin-flip processes when intermediate-energy structures are neglected. Moreover, the larger quadrupolar component at the $K$ edge enhances vibronic excitations, as orbital–lattice entanglement is also time-even in nature.

\begin{figure*}[ht!]
		\centering
		\includegraphics[width=0.99\textwidth]{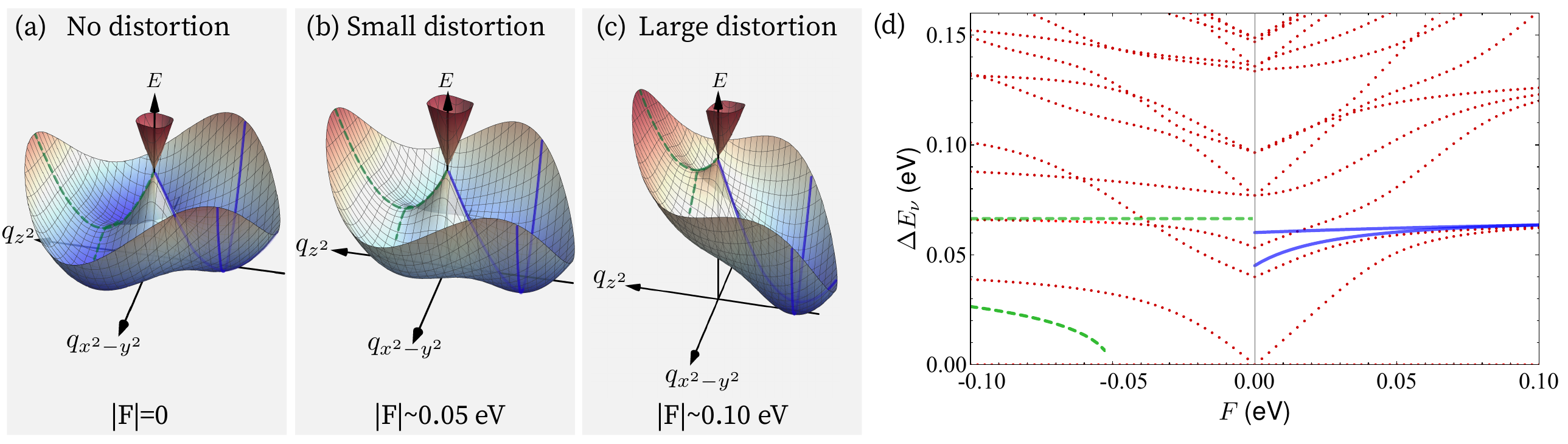}
		\caption{Dynamic JT effect in the elastic field. (a)-(c) The APES's with varied effective elastic field strengths. (d) The evolution of the vibronic energy levels (red dots) with respect to the elastic field arising for tetragonal deformation. The points at $F = 0$ correspond to the vibronic excitation energies shown in Fig.~\ref{Fig:Fig_coupling}(b). $F>0$ ($<0$) corresponds to the tetragonal compression (elongation) in eV. The solid blue and dashed green lines correspond field dependence of the harmonic oscillator frequencies, which are obtained from the parabolic fits depicted on the APES.}
		\label{Fig:Fig2}
	\end{figure*}

    Having discussed how onsite interactions give rise to the vibronic states, we now turn our attention to the intersite interaction between vibronic states on different $d^1$ sites to understand how long-ranged order can develop at low-temperatures. 
    Since the vibronic states [Eq. (\ref{Eq:vibronic})] contain the information on the spin, orbital, and lattice vibrations, they respond to the intersite magnetic exchange, electric quadrupole, and elastic couplings~\cite{Iwahara2023}. Among these various intersite interactions, the elastic coupling is the most dominant and hence is likely the main cause of the cooperative behavior between vibronic states. In the ordered phases, the elastic coupling acts on the local vibronic states as a mean field,
	\begin{align}
		\hat{V}_\text{vib} &= \sum_\gamma \hat{q}_\gamma F_\gamma,
        \label{Eq:Vvib}
	\end{align}
	where $F$'s are the mean elastic fields. The magnitude of $F$ in Ba$_2$CaReO$_6$ is relatively small, on the order of several meV based on temperature scale $T_\mathrm{s}$~\cite{Yamamura2006, Ishikawa2021}.

	The external field alters the shape of the APES and hence modulates the nature of the low-energy vibronic states on sites. For example, under the external field of tetragonal compression type ($F_{z^2} > 0$), one of the minima of the APES is lowered in energy and shifted towards larger deformations [Fig.~\ref{Fig:Fig2}(a)--(c)]. 
    As the magnitude of $F$ increases, the ground vibronic states (denoted by the red dots) split into two Kramers doublets [Fig.~\ref{Fig:Fig2}(d)].
    The distribution of the energy spectrum approaches that of a two-dimensional harmonic oscillator around the global minimum of the APES.
    The solid blue lines in Fig.~\ref{Fig:Fig2}(d) correspond to the two frequencies. 
    Similar behaviour is observed for tetragonal elongation too ($F_{z^2} < 0$), where the two frequencies are denoted by the dashed green lines.

	Next, we consider whether the spin-orbit-lattice entanglement is robust against the elastic field in the case of $5d^1$ DP materials. 
    By using realistic interaction parameters for the rhenium-based DPs of 
    $\lambda = 0.312$ eV, 
    $g = 1.3$, and $\omega = 66$ meV \cite{Pasztorova2023} (see computational details), 
    we find that the ground vibronic state is stabilized by 30 meV, which is more than two times larger than the static JT stabilization energy of 12 meV
    and is three times larger than the effective field of the Re compounds measured by the structural transition temperatures.	Therefore, given that the size of $F$ is very small ($\sim$ 10\,meV), our calculations indicate that the vibronic entanglement effect in cubic $5d^1$ DPs should persist strongly, even below the structural transition temperature. 

\begin{figure}[t!]
		\centering
		\includegraphics[width=0.5\textwidth]{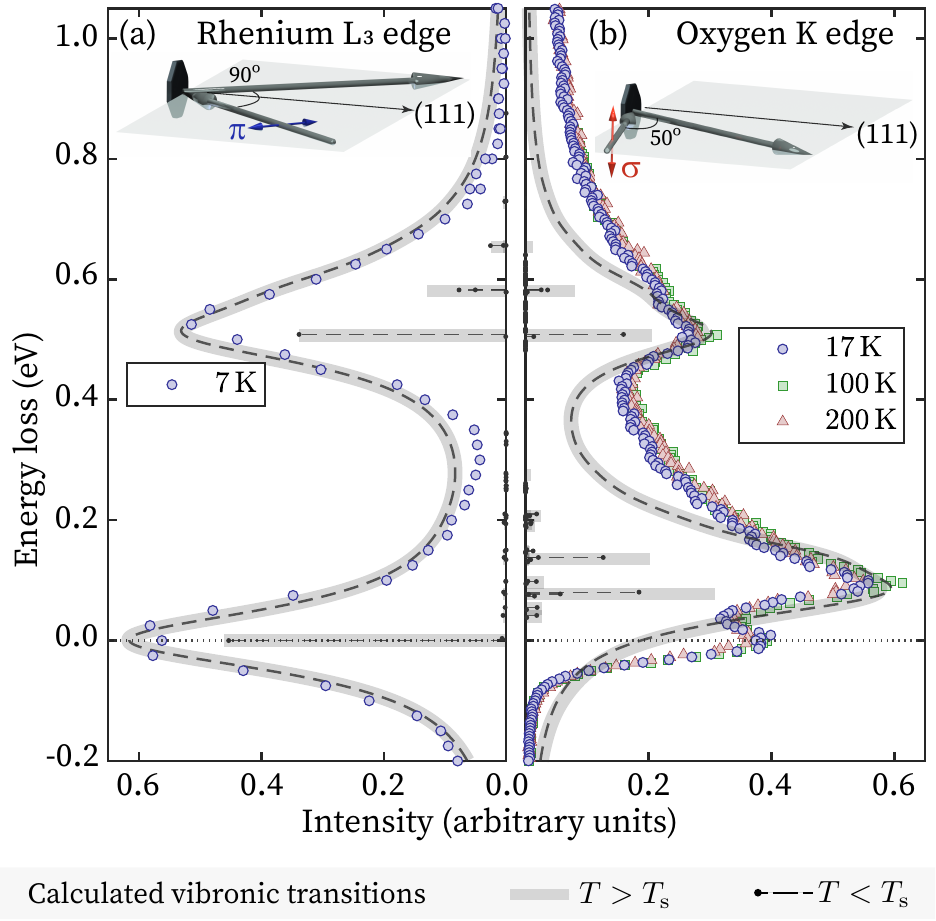}
		\caption{The experimental and calculated low-energy spectra of Ba$_2$CaReO$_6$. (a) The spectra obtained with RIXS at the Re $L_3$-edge at $T$=7 K are denoted by the blue circles. (b) Correspondingly, the experimental spectra collected at the O $K$-edge at $T$= 17, 100, and 200 K, are shown by the $\bigcirc$, $\triangle$, and $\square$ symbols, respectively. In both plots, the calculated transitions between vibronic levels without and with the elastic field are denoted with the thick grey and thin black lines, respectively.  }
		\label{Fig:RIXS}
	\end{figure}

    To experimentally confirm our theoretical prediction, we need to firstly, demonstrate unambiguously the spectroscopic signatures of vibronic states [Fig.~\ref{Fig:Fig_coupling}] and secondly, that these states persist at temperatures well below the structural transition temperature. 
     To search for spectroscopic fingerprints of vibronic dynamics, we performed RIXS measurements on Ba$_2$CaReO$_6$ single crystals at the both Re $L_3$- and O $K$-edge at beamlines BL11XU (SPring-8) and ADRESS~\cite{Strocov,ghiringhelli_saxes_2006} (Swiss Light Source, PSI), respectively. The measured RIXS spectra obtained at various temperatures, above and below $T_\mathrm{s}$, are shown in Fig.~\ref{Fig:RIXS}.
	
The low-energy spectra obtained at the rhenium $L_3$-edge with an incident energy of $E_i$=10.535 keV, displays an inelastic peak at $\Delta E\sim0.5$\,eV [Fig.~\ref{Fig:RIXS}(a)]. Compared to the elastic signal at 0\,eV, this peak at $\sim$0.5\,eV is rather asymmetric with an unusually long tail that extends up to 0.8\,eV in energy transfer, hinting at additional modes between 0.5\,-0.8\,eV. Nonetheless, some of these fine details in the low-energy spectra can be obscured due to the relatively poor energy resolution of RIXS at the Re $L_3$-edge (0.1\,eV). Hence we now turn to consider the low-energy spectra obtained at the oxygen $K$-edge, which has a relatively better energy resolution of $\sim$0.05\,eV [Fig.~\ref{Fig:RIXS}(b)]. In addition to the peak observed at $\Delta E\sim$0.5\,eV, the oxygen RIXS spectra also display another strong inelastic peak at 0.1\,eV, which is distinctly separated from the elastic signal. Both inelastic peaks are markedly asymmetric, with significantly long tails extending up to high-energy transfers. Interestingly, the curves obtained at all three temperatures are very similar, which indicate that the low-energy spectra of Ba$_2$CaReO$_6$ remain unchanged across $T_\mathrm{s}$.
	
 We calculated the Re $L_3$-edge RIXS spectra by adopting the interaction parameters described above and also accounting for the experimental configuration shown in the insert of Fig.~\ref{Fig:RIXS}(a). Our calculations reveal that the vibronic coupling modulates the peak shape of the excited $j_\text{eff}$=$1/2$ multiplet state at 0.52\,eV considerably, resulting in additional transitions at 0.59, 0.67, 0.73 eV. The calculated transition energies and intensities between the vibronic states, denoted by the grey bars in Fig.~\ref{Fig:RIXS}(a), are in good agreement with the experimental data. This correspondence between the two is remarkable given that the calculations were based on an undistorted ReO$_6$ octahedron (i.e. for Ba$_2$CaReO$_6$ above $T_\mathrm{s}$), in contrast to the measurements which were performed at $T$\,=\,7\,K. Hence, for a more representative calculation of the transitions, we have to account for the effect of the structural distortions on the vibronic level, as depicted in Fig.~\ref{Fig:Fig2}.
 
 Curiously, the size of the structural distortion across many $5d^1$ DPs is very subtle, and in some cases remains below the detection limits of x-ray diffraction. This is also true for single crystalline Ba$_2$CaReO$_6$ where the size of the structural distortion is in the order of 0.2\%. Nonetheless, even if we adopt the size of distortion found in polycrystalline Ba$_2$CaReO$_6$~\cite{Yamamura2006, Ishikawa2021} (which are approximately two times larger than those found for the single crystalline samples), we find that the changes to the peak shape are very subtle, resulting only in minor changes to the transition energies and intensities, as denoted by the black lines in Fig.~\ref{Fig:RIXS}(a).

	Similarly, we calculated the oxygen $K$-edge RIXS spectra with the same model as for the $L_3$-edge RIXS, taking into account the experimental configuration shown in the insert of Fig.~\ref{Fig:RIXS}(b). The grey bars indicate the vibronic transitions corresponding to an undistorted ReO$_6$ octahedron at 200 K. Our calculations reveal peaks at 0.52, 0.59, 0.67, 0.73 eV, which arise  due to the 
    vibronic transitions to the $j_\text{eff}=1/2$ manifold with vibrational excitations. 
		
	Our calculation also shows many transitions between the vibronic states from the $j_\text{eff}=3/2$ manifold in the energy range of 0.05-0.4\,eV. 
    As such, these transitions explain that the presence of the inelastic peak at 0.1\,eV in the experimental spectrum [Fig.~\ref{Fig:RIXS}(b)], 
    along with its long tail that extends to high energies, originates from vibronic states: The peak originates from the transitions to the vibronic quartet states at 0.067 eV and 0.104 eV.
    Crucially, the experimental and calculated RIXS spectrum at the oxygen $K$-edge represents direct evidence for the presence of orbital-lattice entanglement in Ba$_2$CaReO$_6$.
		
Next, when the cooperative structural deformation is accounted for, we also find that the distribution of the intensities of the O $K$-edge RIXS spectra does not change considerably. The black lines in Fig.~\ref{Fig:RIXS}(b), denote the calculated transitions at $T$=17\,K, based on the distorted structure described earlier. Our experimental data and calculations both demonstrate clearly that the vibronic states are impervious to the quadrupolar ordering in Ba$_2$CaReO$_6$. Therefore, the orbital-lattice entanglement and, hence, the dynamic JT effect persist.

In summary, we establish that the fine structures in the low-energy spectrum of Ba$_2$CaReO$_6$ are due to the presence of the vibronic states. We demonstrate that these low-energy quantum states strongly persists in the quadrupole ordered phase, despite the structure distortion. As such, our work unambiguously demonstrates that the multipolar phases of the cubic $5d^1$ compounds are characterized by the orbital-lattice entanglement on metal sites. Hence our work opens up a new research field of exotic quantum phenomena driven by the orbital-lattice entanglement.

\begin{acknowledgments}
The authors wish to thank R. Lecamwasam, D. Porter, A. M. Nicholls, and B. Shajilal for their helpful discussions. The theoretical work was supported by a Grant-in-Aid for Scientific Research (Grant No. 22K03507) from the Japan Society for the Promotion of Science, the Iketani Science and Technology Foundation, and the Chiba University Open Recruitment for International Exchange Program. We also acknowledge support from the Singapore National Science Scholarship, Agency for Science Technology and Research and the European Research Council (HERO, Grant No. 810451). The RIXS experiment at the rhenium $L_3$ edge was supported by ``Advanced Research Infrastructure for Materials and Nanotechnology in Japan (ARIM)" of the Ministry of Education, Culture, Sports, Science and Technology (MEXT), Japan (Proposal No. JPMXP1222QS0107) and performed at BL11XU beamline of SPring-8 with the approval of the Japan Synchrotron Radiation Research Institute (JASRI) (Proposal No. 2022B3596). The O $K$-edge RIXS experiments (Proposal No. 20222125)  were performed at the ADRESS beamline of the Swiss Light Source at the Paul Scherrer Institut (PSI). The experimental work at PSI is supported by the Swiss National Science Foundation through project nos. 178867 and 207904. T.Y. is funded by a CROSS project of PSI. Y.W. and G.C.W. acknowledge funding from the European Union’s Horizon 2020 research and innovation programme under the Marie Sklodowska-Curie grant agreement No. 884104 (PSIFELLOW-II-3i program).
\end{acknowledgments}

\appendix

\section{Computational details}
\label{Sec:comput}
We obtained the vibronic states of an isolated $5d^1$ octahedron by numerically diagonalizing the model Hamiltonian consisting of Eqs. (\ref{Eq:HSO}), (\ref{Eq:HJT0}), and (\ref{Eq:HJT}) [See Refs. \cite{Iwahara2018, Frontini2023}].
To this end, we constructed the Hamiltonian matrix with the basis set of $|\Gamma\gamma\rangle \otimes |n_{z^2}, n_{x^2-y^2}\rangle$ ($n_{z^2}, n_{x^2-y^2} = 0, 1, 2, ...$, and $n_{z^2} + n_{x^2-y^2} \le 15$). 
Here, $|n_{z^2}, n_{x^2-y^2}\rangle$ stands for the eigenstates of the two-dimensional harmonic oscillator in Eq. (\ref{Eq:HJT0}). 

To include the mean-field effect on the vibronic states, we numerically diagonalized the above Hamiltonian with Eq. (\ref{Eq:Vvib}).
We determined $F_{z^2}=3.3$ meV so that the expectation value of $\hat{q}_{z^2}$ corresponds to the experimental structure at 17 K \cite{Yamamura2006}.

We consider the RIXS processes involving the transitions between the Re $2p$/O $1s$ to the $5d$ $t_{2g}$ orbitals. 
For the calculations of the RIXS intensities, we employed the Kramers-Heisenberg formula \cite{Sakurai1967, RIXS}. 
\begin{align}
 I(\omega_{k'}) &\propto \sum_{\nu} \rho_\nu \frac{\omega_{k'}}{\omega_k} \left| \sum_\mu \frac{\langle \nu'| \hat{P}_{\bm{k}'\alpha'} | \mu \rangle \langle \mu| \hat{P}_{\bm{k}\alpha} | \nu \rangle}{E_\nu + \hslash \omega_{k} - E_\mu + i \Gamma_\mu}  \right|^2 
 \nonumber\\
 &\times
 \delta \left(E_{\nu'} + \hslash \omega_{k'} - E_\nu - \hslash \omega_{k} \right),
 \label{Eq:KH}
\end{align}
where $\hat{P}_{\bm{k}\alpha}$ is the dipole moments along the polarization vector of the light, 
and $\rho_\nu$ is the canonical distribution for single $5d^1$ site. 
The initial and final states denoted by $\nu, \nu'$ in Eq. (\ref{Eq:KH}) are the vibronic types,
and the intermediate states $\mu$ are 
$5d^2$ vibronic states with a core-hole. 
We used the Hund coupling of 0.26 eV \cite{Yuan2017}.
The broadening of the intermediate states $\Gamma_\mu$ is 
2.25 eV for Re $L_3$-edge \cite{Clancy2012} and 150 meV for O $K$-edge RIXS \cite{johnston_electron-lattice_2016}.
In the O $K$-edge RIXS, the transition occurs 
between the $p$-type molecular orbitals, each of which is composed of two O $1s$ orbitals, and the $t_{2g}$ orbitals.
The $K$ edge scattering amplitude is the same as the sum of the $L_2$ and $L_3$ ones with the degenerate core-hole states. 
To reproduce the RIXS spectra, we convoluted the intensities (\ref{Eq:KH}) with the Lorentzian function: 
The full widths at half maxima are 0.13 eV for Re $L_3$- and 0.1 eV for O $K$-edge RIXS spectra. 
We calculated only the inelastic contributions for the O $K$-edge RIXS spectra. 
We scaled the transition intensities within the $j_\text{eff} = 3/2$ manifold by 0.445 times to account for the energy dependent absorption of low-energy photon. 
In the ordered phases, we averaged the spectra over the domains of the quadrupolar phases.

We determined $\lambda$ and $g$ so that the simulated Re $L_3$-edge RIXS spectra agree with the experimental spectra [Fig. \ref{Fig:RIXS}(a)]. To find the best parameters, we varied the spin-orbit gap by 5 meV and $g$ by 0.05.
The former linearly shifts the position of the $j_\text{eff}=1/2$ peak, and $g$ varies the shape as shown in Fig. \ref{Fig:Fig_coupling}.


%
\end{document}